\documentstyle[preprint,aps]{revtex}

\def\lesssim{\ \raise.3ex\hbox{$<$}\kern-0.8em\lower.7ex\hbox{$\sim$}\ }
\def\gesim{\ \raise.3ex\hbox{$>$}\kern-0.8em\lower.7ex\hbox{$\sim$}\ }
\font\scripti=cmmi7
\font\scriptscripti=cmmi5
\def\sib#1{\setbox0 = \hbox{\scripti #1}
  \kern-.02em\copy0\kern-\wd0
  \kern.04em\box0} % script italic bold 
\def\ssib#1{\setbox0 = \hbox{\scriptscripti #1}
  \kern-.02em\copy0\kern-\wd0
  \kern.04em\box0} % scriptscript italic bold

\font\tenib=cmmib10 % italic bold for math
\skewchar\tenib='177 \skewchar\tenib='177 \skewchar\tenib='177
\def\pbold#1{\setbox0 = \hbox{$ #1 $}
  \kern-.022em\copy0\kern-\wd0
  \kern.011em\copy0\kern-\wd0
  \kern.011em\copy0\kern-\wd0
  \kern.011em\copy0\kern-\wd0
  \kern.011em\box0} % poorman's bold 

\begin{document}
\draft
%\preprint{HEP/123-qed}
\title{The BCS-BEC Crossover in a Gas of Fermi Atoms with a Feshbach Resonance}
\author{Y. Ohashi$^{1,2}$ and A. Griffin$^1$}
\address{$^1$Department of Physics, University of Toronto, Toronto, Ontario, Canada M5S 1A7\\
$^2$Institute of Physics, University of Tsukuba, Ibaraki 305, Japan}
%\date{}
\maketitle
\begin{abstract}
We discuss the BCS--BEC crossover in a degenerate Fermi gas of two hyperfine states interacting close to a Feshbach resonance. This system has quasi-molecular Bosons associated with a Feshbach resonance, and this kind of coupled Boson-Fermion model has been recently applied to Fermi atomic gases within a mean-field approximation. We show that by including fluctuation contributions to the free energy similar to that considered by Nozi\'eres and Schmitt-Rink, the character of the superfluid phase transition continuously changes from the BCS-type to the BEC-type, as the threshold of the quasi-molecular band is lowered. In the BEC regime, the quasi-molecules become stable long-lived composite Bosons, and an effective pairing interaction mediated by the molecules causes pre-formed Cooper-pairs to also be stable even above the transition temperature $T_{\rm c}$. The superfluid phase transition is shown to be usefully interpreted as the Bose condensation of these two kinds of Bosons. The maximum $T_{\rm c}$ is given by (in terms of the Fermi temperature $T_{\rm F}$) $T_{\rm c}=0.218T_{\rm F}$ (uniform gas) and $T_{\rm c}=0.518T_{\rm F}$ (trapped gas in a harmonic potential). 
\par
\vskip3mm
\end{abstract}
\narrowtext
%\newpage
%%%%%%%%%%%%%%%%%%%%%%%%%%%%%%%%%%%%%%%%%%%%%%%%%%%%%%%%%%%%%%%%%%%
%
%\section{Introduction}
The achievement of a superfluid phase transition in a trapped gas of Fermi atoms is one of the most exciting challenges in the current study of ultracold atoms. At present, several groups have produced degenerate Fermi gases down to around 0.2$T_{\rm F}$ (where $T_{\rm F}$ is the Fermi temperature).\cite{Jin1} In the past few years, many theorists have discussed the properties of a superfluid phase based on the BCS concept of Cooper-pair formation.\cite{Stoof,Minguzzi} More recently, this approach has been extended to deal with a Feshbach resonance.\cite{Holland,Timmermans1} A new strongly attractive interaction can arise between Fermions mediated by the quasi-molecular Boson associated with a Feshbach resonance.
However, one must be careful in applying BCS theory to a Fermi gas when the pairing interaction is very strong.  In this case, fluctuations in the two-particle Cooper channel strongly suppress the transition temperature $T_{\rm BCS}$ predicted by weak-coupling BCS theory. Moreover, in this strong coupling regime, one has two ``kinds" of Bosons even in the normal phase above $T_{\rm c}$, i.e., molecules associated with the Feshbach resonance and pre-formed Cooper-pairs. Thus the superfluid phase transition is expected to make a crossover to Bose-Einstein condensation (BEC) from the BCS-like one,\cite{Nozieres,Randeria} which is the subject of the present letter.
\par
Our starting model is identical to that developed in Refs. \cite{Holland,Timmermans1} but we extend their mean-field analysis to include fluctuations in the particle-particle channel. For simplicity, we discuss a uniform Fermi gas. Our work clarifies how the character of the superfluid phase transition gradually changes from the BCS-like transition to BEC-like transition as the threshold energy for the formation of quasi-molecular Bosons is lowered.  
\par
We consider a uniform gas of Fermi atoms with two hyperfine states which interact close to a Feshbach resonance. As usual, we describe these states as pseudo-spins $\sigma=\uparrow,\downarrow$, with $N_\uparrow=N_\downarrow=N/2$. Our model system is described by the Hamiltonian\cite{Holland,Timmermans1}
\begin{eqnarray}
{\hat H}
&=&
\sum_{{\sib p}\sigma}\varepsilon_{\sib p}
c^\dagger_{{\sib p}\sigma}c_{{\sib p}\sigma}
-
U\sum_{{\sib p},{\sib p}'}
c^\dagger_{{\sib p}\uparrow}c^\dagger_{-{\sib p}\downarrow}
c_{-{\sib p}'\downarrow}c_{{\sib p}'\uparrow}
\nonumber
\\
&+&
\sum_{{\sib q}}(E^0_{\sib q}+2\nu)
b^\dagger_{{\sib q}}b_{{\sib q}}
+
g\sum_{{\sib p},{\sib q}}
[
b^\dagger_{{\sib q}}
c_{{\sib p}+{\sib q}/2\uparrow}c_{-{\sib p}+{\sib q}/2\downarrow}
+{\rm h.c}
].
\label{eq.2.1}
\end{eqnarray}
Here $c_{{\sib p}\sigma}$ and $b_{\sib q}$ represent the annihilation operators of a Fermi atom with the kinetic energy $\varepsilon_{\sib p}=p^2/2m$ and a quasi-molecular Boson with the energy spectrum $E^0_{\sib q}+2\nu=q^2/2M+2\nu$, respectively. The threshold energy of the composite Bose particle energy band is denoted by $2\nu$ and is the crucial adjustable parameter in this model. The attractive pairing interaction $-U<0$ originates from non-resonant processes, while $g$ is a coupling constant associated with the Feshbach resonance. Since the $b$-Boson is made of two Fermions, we have $M=2m$ and impose the conservation of the total number of bare Fermi atoms: $\langle {\hat N}\rangle =\langle\sum_{{\sib p}\sigma}c^\dagger_{{\sib p}\sigma}c_{{\sib p}\sigma}\rangle+2\langle\sum_{\sib q}b^\dagger_{{\sib q}}b_{{\sib q}}\rangle$. This crucial constraint is incorporated in the Hamiltonian when we replace ${\hat H}$ by ${\hat H}-\mu {\hat N}$. The resulting ``grand-canonical'' Hamiltonian has the same form as (\ref{eq.2.1}), except that now $\varepsilon_{\sib p}\to\varepsilon_{\sib p}-\mu_{\rm F}$ and $E^0_{\sib q}\to E^0_{\sib q}-\mu_{\rm B}$, with $\mu_{\rm F}=\mu_{\rm B}/2=\mu$. A coupled Fermion-Boson Hamiltonian similar to (\ref{eq.2.1}) has been extensively discussed \cite{Lee} in connection with the high-$T_{\rm c}$ cuprates, usually setting $U=0$.
\par
The superfluid phase transition temperature can be conveniently determined by the Thouless criterion. This describes the instability of the normal phase of Fermions with an attractive interaction associated with the formation of Cooper-pairs of atoms with opposite spins.\cite{Nozieres,Randeria} We first calculate the four-point vertex function $\Gamma(p_+,p_-;p'_+,p_-')$ within a generalized $t$-matrix approximation in terms of $U$ and $g$ (as shown in Fig. 1). One sees that this involves an infinite sum of ladder diagrams involving $U$ and particle-particle bubble diagrams involving $g$. The equation for the particle-particle vertex $\Gamma$ is given by
\begin{eqnarray}
&\Gamma&(p_+,p_-;p'_+,p_-')
=
(g^2D_0(q)-U)
\nonumber
\\
&+&
(g^2D_0(q)-U)T
\sum_{{\sib p}''\omega_m''}
G_0(p''_+)G_0(p''_-)
\Gamma(p''_+,p''_-;p'_+,p_-'),
\label{eq.2.2}
\end{eqnarray}
where $p_+=({\bf p}+{\bf q}/2,i\omega_m+i\nu_n)$, $p_-=({\bf p}-{\bf q}/2,i\omega_m)$ and $q=({\bf q},i\nu_n)$, with $i\omega_m$ and $i\nu_n$ being the Fermion and the Boson Matsubara frequencies, respectively; $G_0^{-1}({\bf p},i\omega_m)=i\omega_m-\varepsilon_{\sib p}+\mu$ and $D_0^{-1}({\bf q},i\nu_n)=i\nu_n-E^0_{\sib q}-2\nu+2\mu$ are, respectively, the bare one-particle Fermion and the Boson thermal Green functions. The solution of (\ref{eq.2.2}) is
\begin{equation}
\Gamma({\bf q},i\nu_n)=-
{U-g^2D_0({\bf q},i\nu_n)
\over
1-[U-g^2D_0({\bf q},i\nu_n)]\Pi({\bf q},i\nu_n)}.
\label{eq.2.3}
\end{equation}
Here $\Pi({\bf q},i\nu_n)$ is the correlation function of the Cooper-pair field operator of two atoms with total momentum ${\bf q}$ defined by ${\hat \Delta}({\bf q})\equiv\sum_{\sib p}c^\dagger_{{\sib p}+{\sib q}/2\uparrow}c^\dagger_{-{\sib p}+{\sib q}/2\downarrow}$, in the absence of $U$ and $g$:\cite{Nozieres,ohashi}
\begin{eqnarray}
\Pi({\bf q},i\nu_n)
=
\sum_{\sib p}
{1-f(\varepsilon_{{\sib p}+{\sib q}/2}-\mu)
-f(\varepsilon_{{\sib p}-{\sib q}/2}-\mu) 
\over 
\varepsilon_{{\sib p}+{\sib q}/2}+\varepsilon_{{\sib p}-{\sib q}/2}-2\mu-i\nu_n
},
\label{eq.2.4}
\end{eqnarray}
where $f(\varepsilon)$ is the Fermi distribution function. $\Pi({\bf q},i\nu_n)$ is often referred to as the particle-particle propagator describing superfluid (Cooper-pair) fluctuations.
\par
 The superfluid phase transition occurs when the particle-particle vertex function in (\ref{eq.2.3}) develops a pole at ${\bf q}=\nu_n=0$. This is easily seen to correspond to the following equation for $T_{\rm c}$:
\begin{equation}
1=(U+g^2{1 \over 2\nu-2\mu})\sum_{\sib p}
{\tanh(\varepsilon_{\sib p}-\mu)/2T_{\rm c} \over 2\varepsilon_{\sib p}-2\mu}.
\label{eq.2.5}
\end{equation}
In (\ref{eq.2.5}), $g^2/(2\nu-2\mu)$ $(=-g^2D_0(0))$ can be interpreted as the additional pairing interaction mediated by a molecular Boson. This interaction becomes very large as $2\mu\to2\nu$. Eq. (\ref{eq.2.5}) defines the temperature $T_{\rm c}$ at which an instability arises in the normal phase of a Fermi gas described by (\ref{eq.2.1}) due to formation of bound states with zero center of mass momentum (${\bf q}=0$) and energy $2\mu$. 
\par
In Refs. \cite{Holland,Timmermans1}, the model in (\ref{eq.2.1}) is solved in the superfluid phase within a simple mean-field approximation in which the $b$-Bosons are replaced by a Bose condensate in the ${\bf q}=0$ state, described by $\langle b_{\sib q=0}\rangle\equiv\phi_m$. Within the BCS-pairing approximation to the two-particle interaction, the Hamiltonian ${\hat H}-\mu{\hat N}$ is then easily diagonalized. One finds the standard self-consistent BCS equation for the effective gap function defined by ${\tilde \Delta}\equiv-Up-g\phi_m$, where $p\equiv\langle\sum_{\sib k}c_{{\sib k}\uparrow}c_{-{\sib k}\downarrow}\rangle$ describes the Cooper-pair Bosons, with the effective attractive interaction $U_{\rm eff}\equiv U+g^2/(2\nu-2\mu)$.
In this mean-field BCS theory below $T_{\rm c}$, the quasiparticle energies are given by $E_{\sib k}=\sqrt{(\varepsilon_{\sib k}-\mu)^2+|{\tilde \Delta}|^2}$. Within this mean-field theory, the condition that the gap function ${\tilde \Delta}$ vanishes can be shown to lead precisely to (\ref{eq.2.5}) as given by the Thouless criterion. Thus (\ref{eq.2.5}) gives a very clear picture of the physics discussed in Refs.\cite{Holland,Timmermans1}.
\par
The chemical potential $\mu$ for our model is determined from the condition that the total number of particles (Fermion+2$\times$Boson) in the interacting system is equal to the number of bare Fermions $N$. As well known, this simply gives $\mu=\varepsilon_{\rm F}$ in the weak-coupling BCS theory if we neglect the temperature dependence of $\mu$ ($\varepsilon_{\rm F}$ is the Fermi energy.). In the mean-field approximation of Refs. \cite{Holland,Timmermans1}, one has simply $N=N_{\rm F}^0+2|\phi_m|^2$ below $T_{\rm c}$. At $T_{\rm c}$, we have $\phi_m=0$. However, $\mu(T_{\rm c}$) in (\ref{eq.2.5}) can be different from $\varepsilon_{\rm F}$ simply because the temperature dependence of $\mu(T)$ in a non-interacting Fermi gas is significant when $T$ is comparable to the Fermi temperature $T_{\rm F}$. It is clear that in Refs. \cite{Holland,Timmermans1}, the transition temperature corresponds to breaking up bound states with ${\bf q}=0$, rather than exciting these molecules into states of finite ${\bf q}$. 
However, $\mu$ can deviate from $\varepsilon_{\rm F}$ in a more fundamental way when the quasi-molecules with ${\bf q}\ne 0$, superfluid fluctuations, and the pre-formed Cooper-pairs are all properly included.\cite{Nozieres,Randeria,Tokumitsu} This is the goal of the present letter. 
\par
The total number of Fermi atoms $N(\mu,T)$ can be obtained from the thermodynamic potential $\Omega$ using the identity $N=-\partial\Omega/\partial\mu$. The effect of the fluctuations in the two-particle vertex function on the potential $\Omega$ is shown in Fig. 2.\cite{Nozieres} In Fig. 2, (a) is the same as the fluctuation free energy diagrams which were studied in Ref.\cite{Nozieres}. In contrast, the diagrams in (b) originate from the Feshbach coupling of the $b$-Bosons and Fermions in (\ref{eq.2.1}), and are new in the present analysis. One finds [$q\equiv({\bf q},i\nu_n)$]
\begin{equation}
N=N^0_{\rm F}+2N^0_{\rm B}-
T\sum_{q}e^{i\delta\nu_n}
{\partial \over \partial\mu}
\ln \Bigl[1-[U-g^2D_0(q)]\Pi(q)\Bigr].
\label{eq.2.6}
\end{equation}
Here $N^0_{\rm F}\equiv2\sum_{\sib p}f(\varepsilon_{\sib p}-\mu)$ and $N^0_{\rm B}\equiv\sum_{\sib q}n_{\rm B}(E^0_{\sib q}+2\nu-2\mu)$, where $n_{\rm B}(E)$ is the Bose distribution function. For $g=U=0$, the $b$-Bosons in (1) start to Bose condensate when $2\nu<2\varepsilon_{\rm F}$ (see Fig. 3(a)). 
The last term in (\ref{eq.2.6}) comes from adding the two kinds of thermodynamic fluctuations shown in Fig. 2. Note that the fluctuations in (\ref{eq.2.6}) are the same as the denominator of the vertex function in (\ref{eq.2.3}). We leave out the convergence factor $e^{i\delta\nu_n}$ in the subsequent frequency sums. 
\par
We must solve (\ref{eq.2.5}) together with (\ref{eq.2.6}) self-consistently. The results we obtain for $\mu(T_{\rm c})$ and $T_{\rm c}$ as a function of the threshold energy $2\nu$ are shown in Figs. 3(a) and 3(b). However, before discussing these numerical results, it is useful to first give a physical interpretation of the contributions in (\ref{eq.2.6}) in terms of the effect of $U$ and $g$ on two kinds of Bose molecules (two-particle bound states). 
We proceed by using the identity $N^0_{\rm B}=-T\sum_{{\sib q},\nu_n}D_0({\bf q},i\nu_n)$ to rewrite (\ref{eq.2.6}) as
\begin{eqnarray}
N&=&N^0_{\rm F}-2T\sum_{q}{\tilde D}(q)
-T\sum_{q}
{\partial \over \partial\mu_{\rm F}}
\ln \Bigl[1-U_{\rm eff}(q)\Pi(q)\Bigr]_{\mu_{\rm F}\to\mu}
\nonumber
\\
&\equiv& N^0_{\rm F}+2N_{\rm B}+2N_{\rm C}.
\label{eq.2.7}
\end{eqnarray}
Here ${\tilde D}^{-1}(q)\equiv i\nu_n-E^0_{\sib q}-2\nu+2\mu+g^2{\tilde \Pi}(q)$ is a renormalized $b$-Boson Green function, with ${\tilde \Pi}(q)\equiv\Pi(q)/[1-U\Pi(q)]$. The second term $2N_{\rm B}$ in (\ref{eq.2.7}) can be interpreted as the contribution of the $b$-Bosons as affected by the self-energy $\Sigma(q)\equiv -g^2{\tilde \Pi(q)}$. The third term $2N_{\rm C}$ in (\ref{eq.2.7}) has the same form as the contribution of fluctuations studied by Nozi\'eres and Schmitt-Rink,\cite{Nozieres} except that now the effective attractive interaction $U_{\rm eff}(q)\equiv U-g^2D_0(q)$ depends on energy as well as momentum. Thus this third term $2N_{\rm C}$ may be viewed as the effect of Cooper-pair fluctuations on $N$. We note that the mean-field approximation discussed in Refs. \cite{Holland,Timmermans1} is reproduced by (\ref{eq.2.7}) by ignoring $2N_{\rm C}$, as well as setting $g^2{\tilde \Pi}(q)=0$ in ${\tilde D}^{-1}$ and only considering the ${\bf q}=0$ $b$-Boson (i.e., setting $E^0_{\sib q}=0$). 
In the strongly coupled Fermion-Boson problem we are dealing with, it should be emphasized that our interpretation of the separate terms in (\ref{eq.2.7}) as $2N_{\rm B}$ and 2$N_{\rm C}$ is only meant as a way of understanding the essential physics. 
\par
In understanding the character of the renormalized $b$-Bosons described by $N_{\rm B}$ in (\ref{eq.2.7}), it is useful to transform the Matsubara frequency summation into a frequency integration,
$
N_{\rm B}=-{1 \over \pi}\sum_{\sib q}\int_{-\infty}^\infty dz
n_{\rm B}(z)
Im{\tilde D}({\bf q},i\nu_n\to z+i\delta)$. 
Here we take the principal value in the $z$-integration. One sees that if the $b$-molecule decays into two Fermi atoms in the presence of the Feshbach resonance, it has a finite lifetime as given by the inverse of the imaginary part of the self-energy $-g^2{\tilde \Pi}$ in ${\tilde D}({\bf q},z+i\delta)$. However, when the chemical potential becomes negative by lowering the threshold energy $2\nu$, a new situation arises: Since it can be shown from ({\ref{eq.2.4}) that $Im\Pi({\bf q},z+i\delta)$ is proportional to the step function $\Theta(z+2\mu-E_{\sib q}^0)$, the renormalized $b$-molecules {\it do not} decay if their energies are smaller than $E^0_{\sib q}-2\mu$. The energy ($\omega_{\sib q}$) of a stable molecule corresponding to the pole of ${\tilde D}$ is given by (for $E^0_{\sib q}-2\mu\ge0$)
\begin{equation}
\omega_{\sib q}=(E^0_{\sib q}-2\mu)+[2\nu-g^2{\tilde \Pi}({\bf q},\omega_{\sib q})].
\label{eq.dd}
\end{equation}
Such stable long-lived molecules appear when the ``effective" threshold $2{\tilde \nu}\equiv2\nu-g^2{\tilde \Pi}({\bf q},z)$ becomes negative, since decay into two Fermi atoms is forbidden. Around $z={\bf q}=0$, we may use the static approximation for this renormalized threshold, $2{\tilde \nu}\simeq 2\nu-g^2{\tilde \Pi}(0,0)$. Using (\ref{eq.2.4}), one can show that ${\tilde \Pi}(0,0)>0$, and hence ${\tilde \Pi}$ is found to promote the generation of stable molecules by decreasing the value of $2{\tilde \nu}$.
\par
$N_{\rm B}$ in (\ref{eq.2.7}) is thus found to consist of two kinds of Bosons, $N_{\rm B}=N_{\rm B}^{\gamma=0}+N_{\rm B}^{\gamma>0}$, i.e., stable molecules $N_{\rm B}^{\gamma=0}$ with infinite lifetime and quasi-molecules $N_{\rm B}^{\gamma>0}$ which can decay into two Fermi atoms (Strictly speaking, $N_{\rm B}^{\gamma>0}$ also includes a contribution from scattering states.). Extracting the contribution of the poles describing these stable molecules, we obtain 
$
N_{\rm B}^{\gamma=0}=\sum_{\bf q}^{\rm poles}Z({\bf q})
n_{\rm B}(\omega_{\sib q})
$, where $Z({\bf q})^{-1}\equiv 1+g^2{\partial {\tilde \Pi}({\bf q},\omega_{\sib q}) \over \partial z}$ describes mass renormalization.
\par
A similar discussion can be used to understand the third term $2N_{\rm C}$ defined in (\ref{eq.2.7}), describing the Cooper-pair fluctuations. When one writes the non-resonant $s$-wave interaction $U$ in terms of the $s$-wave scattering length $U=4\pi aN/m$, one finds $U/\varepsilon_{\rm F}=0.85(p_{\rm F}a)$, where $p_{\rm F}$ is the Fermi momentum. Since a dilute Fermi gas is characterized by $p_{\rm F}a\ll 1$, we must limit ourselves to $U/\varepsilon_{\rm F}\ll 1$. In this case, the non-resonant attractive interaction $U$ in (\ref{eq.2.1}) cannot give rise to the pre-formed Cooper-pairs (in a three-dimensional system).\cite{Nozieres,Randeria,Tokumitsu} However, when $2\mu$ approaches $2\nu$, the interaction $U_{\rm eff}(q)$ mediated by the molecules becomes very strong, so that stable pre-formed Cooper-pairs may appear.\cite{Nozieres} The energy of these poles is found to be precisely the same as that given in (\ref{eq.dd}). Thus we can also divide $N_{\rm C}$ into contributions from (stable) pre-formed Cooper-pairs ($\equiv N^{\gamma=0}_{\rm C}$) and from scattering states ($\equiv N_{\rm C}^{\rm sc}$), with
$
N_{\rm C}^{\gamma=0}={g^2\over 2}
\sum_{\bf q}^{\rm poles}
{\partial {\tilde \Pi}({\bf q},\omega_{\sib q}) \over \partial \mu}
Z({\bf q})n_{\rm B}(\omega_{\sib q})$.
We see that if $g=0$ or if we ignore ${\tilde \Pi}$, there are no stable pre-formed Cooper-pairs and $Z({\bf q})=1$ in $N_{\rm C}^{\gamma=0}$. 

\par
These stable Bosons give a very clear physical picture of BCS--BEC crossover present in the model Hamitonian in (\ref{eq.2.1}).\cite{Holland,Timmermans1} When the threshold $2\nu$ is much larger than $\varepsilon_{\rm F}$, the Fermi states are dominant. In this regime, a BCS-like phase transition is found. Indeed, since $\mu\sim\varepsilon_{\rm F}>0$ in this regime, stable Bosons are absent because $Im{\tilde \Pi}({\bf q},z)\ne 0$ for $z>0$. The only exception is for $z=0$ (since $Im{\tilde \Pi}({\bf q},0)=0$). However, the condition ($2{\tilde \nu}=2\mu$) that a stable Boson with ${\bf q}=0$ appears at $\omega_{\sib q}=0$ given by (\ref{eq.dd}) is easily shown to reduce to the equation for $T_{\rm c}$ in (\ref{eq.2.5}). Namely, as well known in weak-coupling BCS theory, the formation of stable Cooper-pair Bosons and the phase transition occur at the {\it same} temperature.\cite{Nozieres,Randeria} In this limit, no stable long-lived Bosons with ${\bf q}\ne 0$ exist above this transition temperature.
\par
In the opposite limit $\nu\ll 0$, the Fermi states are almost empty. In this regime, the phase transition is expected to be BEC-like. Indeed, since the chemical potential should be negative for $\nu\ll 0$, stable Bosons can appear even above $T_{\rm c}$, as discussed above. The phase transition of the gas of these stable Bosons occurs as usual when the energy of the Boson with ${\bf q}=0$ measured from the chemical potential $2\mu$ reaches zero. Eq. (\ref{eq.dd}) with this condition ($\omega_{\sib q}={\bf q}=0$) again reproduces (\ref{eq.2.5}), i.e., $2\mu=2{\tilde \nu}\equiv2\nu-g^2{\tilde \Pi}(0,0)$ is equivalent to $1=U_{\rm eff}(0)\Pi(0,0)$. In the limit $\nu\ll 0$, the problem has effectively been reduced to the BEC transition in a non-interacting gas of $N/2$-Bosons with mass $M=2m$, with no free Fermions. In this limit, since the superfluid Cooper-pair fluctuations described by $\Pi(q)$ are absent, (\ref{eq.2.5}) [or (\ref{eq.dd}) with $\omega_{\sib q}={\bf q}=0$] simply gives $2\mu=2{\tilde \nu}$, the required condition for a Bose condensate to appear in $N_{\rm B}^{\gamma=0}$ and $N_{\rm C}^{\gamma=0}$. Thus, although (\ref{eq.2.5}) was originally obtained as the condition for the superfluid phase transition in a Fermi gas due to the formation of two-particle bound states, it also describes BEC in a gas mixture of stable $b$-molecules and pre-formed Cooper-pairs ($N_{\rm B}^{\gamma=0}+N_{\rm C}^{\gamma=0}$). 
\par
We now discuss our numerical solution of the coupled equations ($\ref{eq.2.5}$) and (\ref{eq.2.6}), which can be understood in terms of preceding analysis. As usual, a cutoff in calculating $\Pi({\bf q},i\nu_n)$ is needed to remove an (unphysical) ultraviolet divergence. We simply introduce the factor ${\rm exp}[-(\varepsilon/\omega_c)^2]$ in the energy integration, with $\omega_c=2\varepsilon_{\rm F}$. We give energies in units of the bare Fermi energy $\varepsilon_{\rm F}$ and particle number in units of the total number $N$ of bare Fermi atoms. Fig. 3 (a) shows the dependence of $T_{\rm c}$ on the threshold parameter $2\nu$. As $\nu$ is lowered, $T_{\rm c}$ deviates from the BCS limit. At the same time, at $T_{\rm c}$ the character of the particles involved gradually changes from Fermions to composite Bosons. This is shown in Fig. 3(c), as $2\mu$ approaches $2\nu$ in the BEC limit (Fig. 3(b)). We also see in Fig 3(c) that the character of the Bosons also changes from a ``Feshbach-resonance" Boson (quasi-molecule $N_{\rm B}^{\gamma>0}$) to a mixture of stable $b$-molecules and pre-formed Cooper-pairs ($N_{\rm B}^{\gamma=0}+N_{\rm C}^{\gamma=0}$) as $2\nu$ decreases.\cite{foot1} In the limit $\nu\ll 0$, we see that the pre-formed Cooper-pairs disappear, and system reduces to a gas of $b$-Bosons. 
\par
In the BEC limit, since the Fermions, pre-formed Cooper-pairs and superfluid fluctuations are all absent, $T_{\rm c}$ is simply determined by the equation $N=2N^0_{\rm B}$. Using $N=(2mT_{\rm F})^{3/2}/3\pi^2$, we obtain
$T_{\rm c}=2T_{\rm F}
/[6\sqrt{\pi}\zeta(3/2)]^{2/3}
=0.218T_{\rm F}
$.\cite{Nozieres,Randeria} This value of $T_{\rm c}$ gives the {\it maximum} phase transition temperature within our model ($U/\varepsilon_{\rm F}\le 0.3$). 
\par
The BCS--BEC crossover induced by changing the strength of the pairing interaction has been investigated as a model for high-$T_{\rm c}$ superconductors.\cite{Randeria,Tokumitsu} In the present case, when we plot $T_{\rm c}$ with respect to the effective pairing interaction, $U_{\rm eff}=U+g^2/(2\nu-2\mu)$, we obtain the inset in Fig. 3(b). This is analogous to results in Ref. \cite{Randeria}. Since $2\mu$ approaches $2\nu$ as $\nu$ decreases (see Fig. 3(b)), the strong coupling regime in the inset corresponds to the case of small $\nu$ in Fig. 3(a). 
\par
For trapped Fermions, one expects that the {\it maximum} $T_{\rm c}$ will again be given by the BEC transition for $N/2$-Bosons but now in a parabolic trap. This gives $T_{\rm c}=T_{\rm F}(1/6\zeta(3))^{1/3}=0.518T_{\rm F}$, where $T_{\rm F}=(3N)^{1/3}\omega_0$.
\par
To summarize, we have generalized the analysis in Refs. \cite{Holland,Timmermans1} to show that the superfluid phase transition in a uniform gas of Fermi atoms with a Feshbach resonance can change from the BCS-like to BEC-like when we include the fluctuations associated with the Cooper-pair (or two-particle) channel. We have also introduced a physical interpretation of our results in terms of two kinds of Bose molecules. In contrast with superconductors, the BCS--BEC crossover in atomic Fermi gases appears to be accessible by lowering the threshold energy $2\nu$ of the quasi-molecular Bosons associated with the Feshbach resonance. Generalization of these results to below $T_{\rm c}$ and for a trapped Fermi gas within the local density approximation (LDA) will be discussed elsewhere.
\par
{\it Acknowledgements.} We thank Professor S. Takada, Dr. T. Nikuni and Dr. J. Ranninger for useful discussions. Y. O. acknowledges a Japanese Overseas Research Fellowship for support at the University of Toronto. A.G. acknowledges support from NSERC of Canada.
%
%%%%%%%%%%%%%%%%%%%%%%%%%%%%%%%%%%%%%%%%%%%%%%%%%%%%%%%%%%%%%%%%%%%%%
%
%%%%%%%%%%%%%%%%%%%%%%%%%%%%%%%%%%%%%%%%%%%%%%%%%%%%%%%%%%%%%%%%%%%%

%%%%%%%%%%%%%%%%%%%%%%%%%%%%%%%%%%%%%%%%%%%%%%%%%%%%%%%%%%%%%%%%%%%%%%%%%%%%%%
%
\begin{figure}
\caption{Generalized $t$-matrix approximation for the two-particle vertex function $\Gamma$. The solid (dashed) line represents $G_0$ ($D_0$), the dotted line describes $-U$, and the vertex $g$ is the coupling parameter defined in (1).}
\label{Fig1} 
\end{figure}
\begin{figure}
\caption{Thermodynamic potential $\Omega$ fluctuations associated with the two-particle Cooper-pair channel. Infinite series of these kinds of diagrams are summed.}
\label{Fig2}
\end{figure}

\begin{figure}
\caption{Self-consistent solution for (a) $T_{\rm c}$ and (b) $\mu(T_{\rm c})$ as a function of $\nu$, $U=0.3\varepsilon_{\rm F}$, $g=0.6\varepsilon_{\rm F}$, and a cutoff $\omega_c=2\varepsilon_{\rm F}$. In panel (a), for comparison, we show the BEC case ($g=U=0$), and BCS solution [4] of (5) with $\mu(T)$ being the chemical potential of a non-interacting Fermi gas. Panel (c) shows the number of Fermions $N_{\rm F}^0$, stable $b$-molecules $N_{\rm B}^{\gamma=0}$, quasi-molecules (finite lifetime) $N_{\rm B}^{\gamma>0}$, pre-formed Cooper-pairs $N_{\rm C}^{\gamma=0}$, and the scattering part $N_{\rm C}^{\rm sc}$, all at $T_{\rm c}$, as a function of the Feshbach threshold parameter $2\nu$. In the case shown, stable long-lived Bosons only exist for $\nu\lesssim 0.22 \varepsilon_{\rm F}$.} 
\label{Fig3}
\end{figure}
%%%%%%%%%%%%%%%%%%%%%%%%%%%%%%%%%%%%%%%%%%%%%%%%%%%%%%%%%%%%%%%%%%%
%
%
%

\begin{references}
\bibitem{Jin1} B. DeMarco and D. S. Jin, Science {\bf 285}, 1703 (1999); A. G. Truscott et al., Science {\bf 291}, 2570 (2001); F. Schreck et al., Phys. Rev. Lett. {\bf 87}, 080403 (2001).
\bibitem{Stoof} For a review, see H. T. C. Stoof and M. Houbiers, {\it Bose-Einstein Condensation in Atomic Gases}, ed. M. Inguscio, S. Stringari and C. E. Wieman (IOS press, Amsterdam, 1999), p.537.
\bibitem{Minguzzi} A. Minguzzi, G. Ferrari and Y. Castin, cond-mat/0103591. Further references are given here.
\bibitem{Holland} M. Holland et al., Phys. Rev. Lett. {\bf 87}, 120406 (2001). This work has been extended to a trapped gas in M. L. Chiofalo et al., cond-mat/0110119.

\bibitem{Timmermans1} E. Timmermans et al., Phys. Lett. A {\bf 285}, 228 (2001); E. Timmermans et al., Phys. Rep. {\bf 315}, 199 (1999).
\bibitem{Nozieres} P. Nozi\'eres and S. Schmitt-Rink, J. Low Temp. Phys. {\bf 59}, 195 (1985).
\bibitem{Randeria} For a review, see M. Randeria, {\it Bose-Einstein Condensation}, ed. A. Griffin, D. W. Snoke and S. Stringari (Cambridge, N.Y., 1995), p.355.
\bibitem{Lee} R. Friedberg and T. D. Lee, Phys. Rev. B. {\bf 40}, 6745 (1989); J. Ranninger, {\it Bose-Einstein Condensation}, ed. A. Griffin, D. W. Snoke and S. Stringari (Cambridge, N.Y., 1995), p.393. References to the earlier literature are given here ; T. Kostyrko and J. Ranninger, Phys. Rev. B {\bf 54}, 13105 (1996).
\bibitem{ohashi} Y. Ohashi and S. Takada, J. Phys. Soc. Jpn. {\bf 66}, 2437 (1997). 
\bibitem{Tokumitsu} A. Tokumitsu, K. Miyake and K. Yamada, Phys. Rev. B {\bf 47}, 11988 (1993).
\bibitem{foot1} In Fig. 3(c), the number of Fermions $N_{\rm F}^0$ does not reach $N$ even for $\nu\gesim2\varepsilon_{\rm F}$, where the effect of the $b$-Bosons is negligible. This is because the scattering part $N_{\rm C}^{\rm sc}$ in the last term in (\ref{eq.2.7}) still makes a contribution to $N$, as shown in Fig. 3(c). This scattering contribution $N_{\rm C}^{\rm sc}$ is negative for $\nu\lesssim 0.14\varepsilon_{\rm F}$. In this regard, we note that there are cancellations between the various contributions in (7) and it is only the sum which gives $N(\mu,T)$.
%\bibitem{Inouye} See, for example, S. Inouye, M. R. Andrews, 
%J. Stenger, H. -J. Miesner, D. M. Stamper-Kurn and W. Ketterle, 
%Nature (London) {\bf 392},151 (1998); Ph. Courteille, R. S. 
%Fregland, D. J. Heizen, F. A. van Abeelen and B. J. Verhaar, 
%Phys. Rev. Lett. {\bf 81}, 69 (1998); S. L. Cornish, 
%N. R. Claussen, J. L. Roberts, E. A. Cornell and C. E. Wieman, 
%Phys. Rev. Lett. {\bf 85}, 1795 (2000).

\end{references}
\end{document}